\documentclass[aps,floats,prc,showpacs]{revtex4}
\usepackage{graphicx}
\usepackage[dvips]{color}
\usepackage{colordvi}
\def\Journal#1#2#3#4{{\em #1} {\bf #2}, #3 (#4) }
\def\NPA{{ Nucl. Phys.} A}

\def\PRL{Phys. Rev. Lett.}

\def\PRC{{Phys. Rev.} C}
\def\PL{Phys. Lett.}
\def\PLB{Phys. Lett. B}

\begin{document}
\title{Correlations and spectral functions in asymmetric nuclear matter}
\author{Kh.S.A. Hassaneen  and H. M\"uther}
\affiliation{Institut f\"ur
Theoretische Physik, \\ Universit\"at T\"ubingen, D-72076 T\"ubingen, Germany}
\begin{abstract}
The self-energy of nucleons in asymmetric nuclear matter is evaluated employing
different realistic models for the nucleon-nucleon interaction. Starting from
the Brueckner Hartree Fock approximation without the usual angle-average in the
two-nucleon propagator the effects of the hole-hole contributions are
investigated within the self-consistent Green's function approach. Special
attention is paid to the isospin-dependence of correlations, which can be
deduced from the spectral functions of nucleons in asymmetric matter. The strong
components of the proton - neutron interaction lead in neutron-rich matter
to a larger depletion for the occupation probability of proton states below the
Fermi momentum.   
\end{abstract}
\maketitle

\section{Introduction\label{Introduction}}
The development of the nuclear shell model has been one of the keys for a
microscopic understanding of nuclear structure. Therefore a lot of effort has
been made to deduce the properties of the mean field of nucleons in nuclei from
realistic models of the nucleon-nucleon (NN) interaction. The simple
Hartree-Fock approach fails badly if one employs realistic NN interactions, i.e.
interactions which are adjusted to describe the NN scattering data: such Hartree
Fock calculations typically yield unbound nuclei and no binding energy for
nuclear matter at saturation density.

This demonstrates the necessity to account for the effects of correlations,
which are induced from the strong short-range components and the tensor
components of realistic NN forces. Therefore various methods for treating
strongly interacting many Fermion systems have been developed and applied to
account for these correlation effects (see e.g.\cite{bookbal} and \cite{mupo}
for a review on this topic). One of the most popular approaches in nuclear
physics has been the hole-line expansion and in particular the lowest order
approach in this expansion, the Brueckner-Hartree-Fock (BHF) approximation. 

Like the ordinary Hartree-Fock approach also the BHF approximation assumes a
Slater-determinant for the nuclear wave function and the effects of correlations
are taken into account by means of an effective interaction, the so-called $G$
matrix. This $G$ matrix is obtained from solving the Bethe-Goldstone equation,
which in some sense corresponds to the NN scattering $T$ matrix. In contrast to
the Lippmann-Schwinger equation, leading to $T$, the Bethe-Goldstone equation
accounts for effects of the nuclear medium: The propagator for the intermediate
two-particle states is restricted to particle states, i.e. to single-particle
states with energies above the Fermi energy $\varepsilon_F$, and is defined in
terms of the single-particle energies for the nucleons in the medium. Therefore
the solution of the Bethe-Goldstone equation, the $G$ matrix accounts for
particle-particle ladders. The single-particle energies are determined from the
Hartree-Fock approximation for the nucleon self-energy or single-particle
potential, however, replacing the bare NN interaction by the effective
interaction, the $G$ matrix. This $G$ matrix is also used to determine the total
energy of the system.

The BHF approximation may also be considered as a first step towards a
many-body approach, which is based on a a self-consistent evaluation of Green's
function\cite{dickbarb}.   Within the scheme of self-consistent Green's
function, however, the evaluation of the self-energy is done employing a
scattering matrix for two nucleons in the medium, which accounts for
particle-particle as well as hole-hole ladders. In nuclear physics typical
values for the Fermi momentum are small compared to the single-particle
momenta which can be reached by NN scattering processes employing realistic NN
forces. Therefore the contributions of the particle-particle ladder terms are
more important than those of the hole-hole terms, which provides a
justification for the approximation used in the hole line  expansion.

Therefore the main advantage of the self-consistent Green's function approach is
not the inclusion of hole-hole terms in the effective NN scattering matrix but
that it yields a consistent result for the single-particle Green function, which
can be analyzed to determine the spectral functions of nucleons in the nuclear
medium. These spectral functions directly reflect the effects of the NN
correlations and can be explored by the analysis of nucleon knock-out
experiments like $(e,e'p)$.

Various attempts have been made to perform self-consistent calculations of the
Green's function for nucleons in nuclear matter using realistic NN forces. A
serious problem of those investigations is the appearance of the so-called
pairing  instability\cite{vonder,alm,bozek}.  Such pairing effects can be taken
into account by means of the BCS approach\cite{baldbcs,almbcs,elgar}. At the
empirical saturation density of symmetric nuclear matter the solution of the
gap equation in the $^3S_1-^3D_1$ partial wave leads to an energy gap of around
10 MeV. Another approach is to consider an evaluation of the generalized ladder
diagrams with ``dressed'' single-particle propagators. This means that the
single-particle Green's functions are not approximated by a mean-field approach
but consider single-particle strength distributed over all energies. Various
attempts have been made in this direction, considering a parameterization of
the single-particle Green's function in terms of various poles\cite{dimitr},
employing simplified (separable) interaction models\cite{bozek} or considering
the case of finite temperature\cite{fricn}.
                                                                                
It is worth noting that the same instabilities have also been observed in
studies  of finite nuclei\cite{heinz}, leading to divergent contributions to
the binding  energy from the generalized ring diagrams. These contributions
remain finite if the single-particle propagator are dressed in a
self-consistent way.

An approximative scheme for the self-consistent evaluation of the
single-particle Green's function has been developed and applied to symmetric
nuclear matter in \cite{khaga1}. In this approach the $G$ matrix is evaluated
avoiding the angle-average approximation for two-nucleon propagator which is
usually considered. The effects of the hole-hole scattering terms are taken into
account by means of a perturbative approach and the single-particle propagator
are approximated by a single-pole approach, deriving the energy of the pole from
the energy weighted spectral distribution. 

It has been shown that this approach
yields results for the spectral function, which are in  good agreement
with a determination in a complete self-consistent Green's function approach.
However, it underestimates the hole strength at very high momenta and missing
energies\cite{rohe}.

In this manuscript we present an approximation scheme close to the one
developed in \cite{khaga1} and apply it to asymmetric nuclear matter. The main
aim of this study is to explore the dependence of the spectral functions for
protons and neutrons on  the nuclear asymmetry and relate this to the
correlations induced by the proton-proton (neutron-neutron) and proton-neutron
interaction. Since the tensor components of the NN interaction are suppressed
in the proton-proton and neutron-neutron interaction, this analysis shall also
allow us to distinguish the correlation effects originating from the tensor
components from those induced by the central short-range components of the NN
interaction. 

The
dependence of these results on the model for the NN interaction is
explored by comparing the results derived from two modern NN interactions, which
fit NN phase shifts with high accuracy. One of them, the so-called Nijm2
interaction, is a localized version of the Nijmegen interaction\cite{nijm1}. The
other one, the CD-Bonn potential, is defined in momentum space and no attempt is
made to localize it\cite{cdb}. 

After this introduction we will shortly review the scheme to evaluate the
spectral function within the Green's function approach. Results will be
presented in section 3 including a discussion of the usual angle-average
approximations for the NN propagator in the scattering equation. The main
conclusions are summarized in section 4.

\section{Self-energy and single-particle Green's function}

The single-particle potential or self-energy of a nucleon with isospin $\tau$,
momentum $\vec k$ and energy $\omega$ in asymmetric nuclear matter is defined
in the Brueckner-Hartree-Fock (BHF) approximation by
\begin{equation}
\Sigma^{BHF}_{\tau} (\vec k,\omega ) = \sum_{\tau'} \int d^3q <\vec k\vec q |
G(\Omega ) |\vec k \vec q >_{\tau\tau'} n^0_{\tau'}(\vec q)
\,.\label{eq:selfbhf}
\end{equation}
Note that here and in the following we suppress the spin quantum numbers in
order to simplify the notation. In this equation $n^0_{\tau}(\vec q)$ refers to
the occupation probability of a free Fermi gas of protons ($\tau = \pi$) and
neutrons ($\tau = \nu$). This means for asymmetric matter with a total density
$\rho$ and asymmetry $\alpha$
\begin{eqnarray}
\rho & = & \rho_{\pi} + \rho_{\nu} \nonumber\\
\alpha & = & \frac{\rho_{\nu}-\rho_{\pi}}{\rho}\,,\label{def:rhobet}
\end{eqnarray}
this occupation probability is defined by
\begin{equation}
n^0_{\tau} (\vec q) = \left\{ \begin{array}{ll} 1 & \mbox{for}\; |\vec q| \leq
k_{F\tau} \\
0 & \mbox{for}\; |\vec q| > k_{F\tau} \end{array}\right.\label{eq:occ0}
\end{equation}
with Fermi momenta for protons ($k_{F\pi}$) and neutrons ($k_{F\nu}$) which are
related to the corresponding densities by
\begin{equation}
\rho_{\tau} = \frac{1}{3\pi^2} k_{F\tau}^3\,.\label{eq:kftau}
\end{equation}
The matrix elements in (\ref{eq:selfbhf}) refer to the anti-symmetrized $G$
matrix elements, which for a given NN interaction $V$ are obtained by solving
the Bethe-Goldstone equation
\begin{eqnarray}
<\vec k\vec q | G(\Omega ) |\vec k\vec q >_{\tau\tau'} & = 
<\vec k\vec q | V |\vec k\vec q > + & \int d^3p_1 d^3p_2 <\vec k\vec q | V
|\vec p_1 \vec p_2 >_{\tau\tau'}\nonumber \\ && \times
\frac{Q(p_1\tau,p_2\tau')}{\Omega - (\varepsilon_{p1,\tau} +
\varepsilon_{p2\tau'}) + i\eta} <\vec p_1 \vec p_2| G(\Omega ) |\vec
k\vec q >_{\tau\tau'} \,.\label{eq:betheg}
\end{eqnarray}
The single-particle energies $\varepsilon_{p\tau}$ should be identified with the
BHF single-particle energies which are defined in terms of the real part of the
BHF self-energy of (\ref{eq:selfbhf}) by
\begin{equation}
\varepsilon_{k\tau} = \frac{k^2}{2m} +  \mbox{Re}\left[ \Sigma^{BHF}_{\tau} (
\vec k,\omega = \varepsilon_{k\tau})\right]
\,,\label{eq:bhf1}
\end{equation}
with a value for the starting energy parameter $\Omega$ in the Bethe-Goldstone
equation (\ref{eq:betheg}) of
\begin{equation}
\Omega = \omega + \varepsilon_{q\tau'} = \varepsilon_{k\tau} +
\varepsilon_{q\tau'}\,.
\label{eq:bhf2}
\end{equation}
The Pauli operator $Q(p_1\tau,p_2\tau')$ in the Bethe Goldstone
Eq.(\ref{eq:betheg}) is used to restrict the intermediate states to
particle-particle states, i.e. to states in which the momenta of nucleons with
isospin $\tau$ are above the corresponding Fermi momentum. This Pauli operator
as well as the single-particle energies for the intermediate states are defined
in terms of the single-particle momenta $p_1$ and $p_2$. Often, however, one
uses a parameterization of the single-particle spectra in terms of an effective
mass in the form 
\begin{equation}
\varepsilon_{k\tau} \approx \frac{k^2}{2m^*_{\tau}} + C_{\tau} \,. \label{eq:param1}
\end{equation}
This parameterization allows the definition of a so-called angle-averaged
propagator, which reduces the Bethe-Goldstone equation to an integral equation
in one dimension, to be solved for various partial waves. In our calculations
we avoid this angle average procedure and solve the Bethe Goldstone equation
with the exact propagator using the techniques discussed in \cite{schiller}.

In a next step we extend the definition of the self-energy and include the
effects of hole-hole scattering terms in a kind of perturbative way\cite{gcl}
\begin{equation}
\Delta \Sigma^{2h1p}_{\tau} (k,\omega) = \sum_{\tau'}\int_{k_{F\tau'}}^{\infty} 
d^3p \int_0^{k_{F\tau}} d^3h_1\,\int_0^{k_{F\tau'}}d^3h_2\, 
\frac{<k,p\vert G\vert h_1,h_2>_{\tau\tau'}^2}{\omega +
\varepsilon_{p\tau'} - \varepsilon_{h_1\tau} -
\varepsilon_{h_2\tau'}-i\eta}\,.\label{eq:2h1p}
\end{equation}
The comparison of the contributions from particle-particle ladders, contained in
$\Sigma^{BHF}$, and those originating from the hole-hole ladders for symmetric
nuclear matter in \cite{khaga1} justifies the perturbative treatment of 
$\Delta \Sigma^{2h1p}$. Also for this extended definition of the self-energy we
can define a quasi-particle energy by
\begin{equation}
\varepsilon_{k\tau}^{\mbox{QP}} = \frac{k^2}{2m} +  \mbox{Re}\left[ 
\Sigma^{BHF}_{\tau} (\vec k,\omega = \varepsilon_{k\tau}^{\mbox{QP}}) +
\Delta \Sigma^{2h1p}_{\tau}  (\vec k,\omega = \varepsilon_{k\tau}^{\mbox{QP}})
\right]
\,.\label{eq:qpbhf}
\end{equation}
The quasi-particle energy for the momentum $k=k_{F\tau}$ defines the Fermi
energy for protons and neutrons, respectively
\begin{equation}
\varepsilon_{F\tau} = \varepsilon_{k_F\tau}^{\mbox{QP}}\,.\label{eq:fermiqp}
\end{equation}
The real part (Re) and imaginary part (Im) of the self-energy, $\Sigma =
\Sigma^{BHF}+ \Delta \Sigma^{2h1p}$ can also be  used to determine the spectral
functions $S^h_{\tau}(k,\omega)$ and $S^p_{\tau}(k,\omega)$ for hole and
particle strength, respectively
\begin{equation}
S^{h(p)}_{\tau}(k,\omega)= \pm \frac {1}{\pi} \frac {\mbox{Im}
\Sigma_{\tau}(k,\omega)} {(\omega -k^2/2m - \mbox{Re} \Sigma_{\tau}
(k,\omega))^2 + (\mbox{Im} \Sigma_{\tau} (k, \omega))^2} ~,~ \mbox{for} \
\omega < \varepsilon_{F\tau} ~ (\omega > \varepsilon_{F\tau})\,. 
\label{eq:spectf}
\end{equation} 
The hole spectral function represents the probability that a particle with
isospin $\tau$, momentum $k$ and energy $\omega$ can be removed from the 
ground state of the system, leaving the residual nucleus with $(A-1)$ nucleons
in an eigenstate  of the Hamiltonian with an energy $E_{A-1} = E_{A0}-\omega$,
where $E_{A0}$ refers to the ground-state energy of the original system.
Integrating the spectral distribution of the hole states yields the occupation
probability
\begin{equation}
n_{\tau}(k) =  \int_{-\infty}^{\varepsilon_{F\tau}} d\omega\,S^h_{\tau}
(k,\omega)\,.\label{eq:n(k)}
\end{equation}
The mean energy for the distribution of the hole strength is obtained by
\begin{equation}
\langle\varepsilon_{h\tau}(k)\rangle = \frac{\int_{-\infty}^{\varepsilon_{F\tau}} 
d\omega\,\omega\,S^h_{\tau}(k,\omega)}{n_{\tau}(k)}\,,\label{eq:meaneh}
\end{equation}
and the corresponding energy for the distribution of the particle strength can
be calculated as
\begin{equation}
\langle\varepsilon_{p\tau}(k)\rangle = \frac{\int^{\infty}_{\varepsilon_{F\tau}} 
d\omega\,\omega\,S^p_{\tau}(k,\omega)}{1-n_{\tau}(k)}\,.\label{eq:meanep}
\end{equation}
Our approximation to a self-consistent Green's function (GF) calculation can
now be defined by identifying the single-particle to be used in the energy
denominators of  the Bethe-Goldstone equation (\ref{eq:betheg}) as well as in
the calculation of the 2h1p correction term in (\ref{eq:2h1p}) by
\begin{equation}
\varepsilon_{k\tau} = \left\{\begin{array}{ll}
\langle\varepsilon_{h\tau}(k)\rangle &
\mbox{for $k<k_{F\tau}$}\\ \langle{\varepsilon_{p\tau}}(k)\rangle &
\mbox{for $k>k_{F\tau}$}\end{array}\right.\,.\label{eq:selfeps}
\end{equation}
With this definition we employ a single-particle Green's function, which is
defined for each momentum $k$ by just one pole at the energy $\omega = 
\varepsilon_{k\tau}$. This means that in the calculation of the self-energy we
account for the fact that the spectral strength is distributed over a wide
range of energies and consider the mean value of the distribution. This leads to
an energy spectrum exhibiting a small gap at $k=k_F$. This gap is of similar
size as the pairing gap obtained in BCS calculations. With this choice we
circumvent the problems of the pairing instabilities. However, in calculating
the terms of the self-energy we
do not consider a depletion of the occupation for momenta $k$ below the Fermi
momentum and partial occupation of states with momenta $k$ larger than $k_F$.

This partial depletion and occupation is taken into account, when the total
energy per nucleon is determined by
\begin{equation}
\frac{E}{A} = \frac{\sum_{\tau}\int d^3k\,\int_{-\infty}^ {\varepsilon_{F\tau}}
 d\omega\,
S^h_{\tau}(k,\omega)\frac{1}{2}\left(\frac{k^2}{2m} + \omega\right)}
{\sum_{\tau}\int d^3k\,n_{\tau}(k)}
\,.\label{eq:ebhf2}
\end{equation}

\begin{figure}
\begin{center}
\center\includegraphics[scale=0.45]{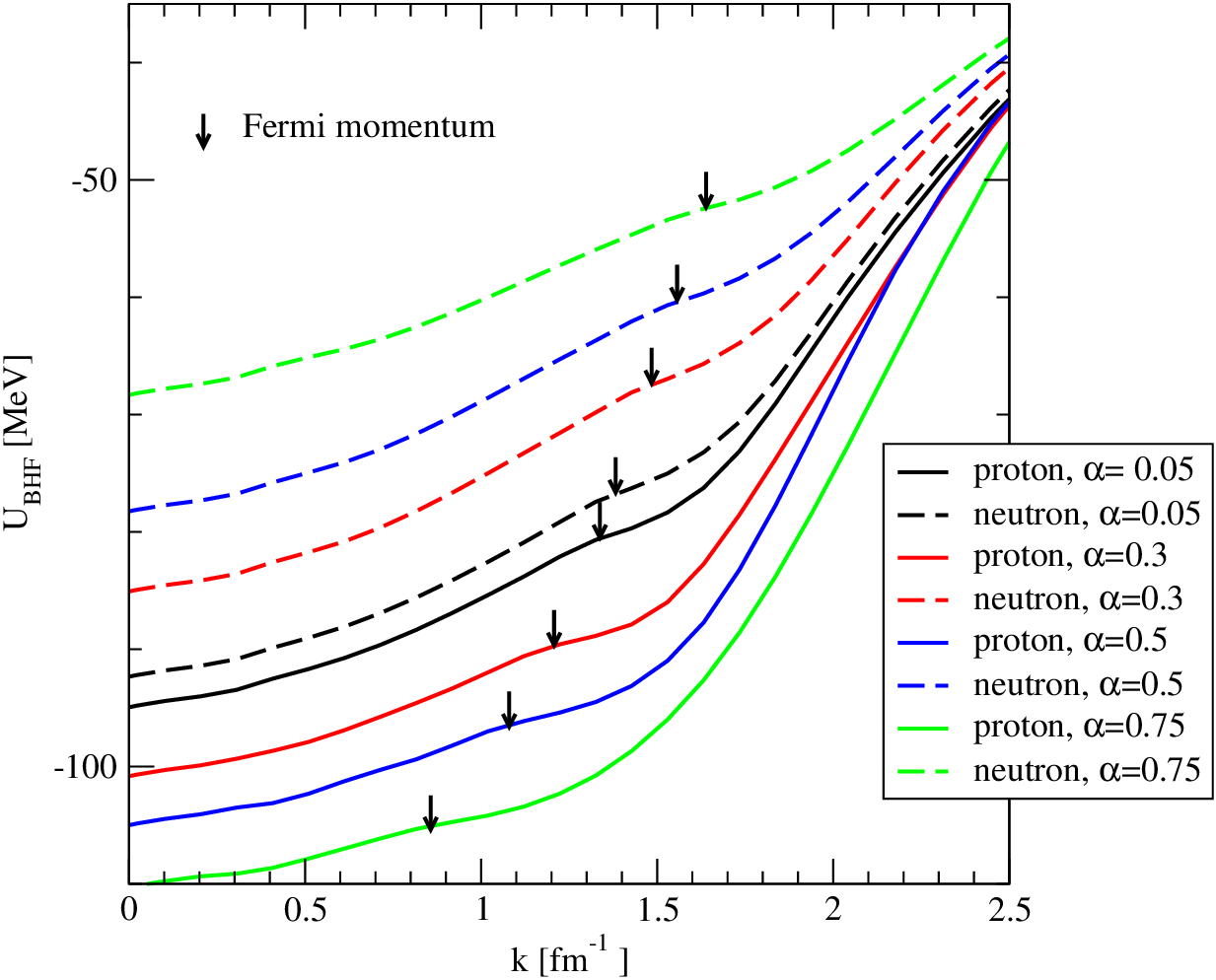}
\end{center}
\caption{\label{fig:epsi1}(Color online) The self-energy for protons (solid
lines) and neutrons (dashed lines) in the BHF approximation. Results were
obtained using the CD-Bonn interaction at density $\rho$ = 0.17 fm$^{-1}$ for
various values of the asymmetry parameter $\alpha$.}
\end{figure}

\section{Results and Discussion}
 
As a first step we would like to discuss the self-energies or single-particle
potentials as a function of the asymmetry of nuclear matter. For that purpose we
consider nuclear matter at the empirical value for the saturation density of
symmetric nuclear matter ($\rho$ = 0.17 fm$^{-1}$) and consider the
self-energies for various values of the asymmetry parameter $\alpha$. 

Results for the BHF approximation of the self-energy for protons and neutrons
are displayed in Fig.~\ref{fig:epsi1} for values of the asymmetry parameter
$\alpha$ ranging from 0.05 to 0.75. The CD-Bonn interaction has been used to
generate the results for this figure. 

One finds that the single-particle potentials for protons are more attractive
than those for neutrons and that the depth of the potential for the protons,
the value of the self-energy at $k=0$ is more attractive with increasing
asymmetry. This reflects the fact that the effective interaction is more
attractive between protons and neutrons than between nucleons of the same
isospin. For the CD-Bonn interaction this feature can be observed to some
extent already on the level of the Hartree-Fock (HF) approximation. The HF
self-energies, however, are less attractive and the dependence of the depth of
the self-energies is weaker. For $\alpha=0.75$ the  values for the HF
self-energies at $k=0$ range from -52 MeV to -42 MeV for protons and neutrons,
respectively. Using the local interaction Nijm2 of the Nijmegen group one
obtains results in the BHF approximation, which are rather similar to those
displayed in Fig.~\ref{fig:epsi1}. Using the HF approximation, however, the
Nijm2 interaction yields much less attractive self energies, ranging from 52
MeV to -8 MeV and in this case the self-energy is  more attractive for the
neutrons than for the protons.

This is already a first indication for the differences between the two
interactions considered: The local interaction Nijm2 is stiffer than the
non-local CD-Bonn potential. Therefore a larger part of the attraction in 
the effective interaction originates from the particle-particle ladder
contributions to the $G$-matrix. This is true in particular for the
proton-neutron interaction, which can be traced back to the correlations in the
$^3S_1 - ^3D_1$ channel of the interaction of nucleons with total isospin 
$T=0$\cite{local}.

From Fig.~\ref{fig:epsi1} one can also see that the BHF self-energies do not
have a simple parabolic shape as a function of the momentum. There is a
characteristic {\em dip} in the self-energy always occurring at momenta slightly
above the Fermi momentum of the kind of nucleons under consideration. The
momentum dependence of the single-particle potential in a homogeneous infinite 
system is a sign of the non-locality of the single-particle potential and is 
typically characterized in terms of an effective mass $m^*$, which can be used
to parameterize the momentum dependence of the single-particle energies
according Eq.~(\ref{eq:param1}). This non-locality of the single-particle 
potential can be due to a non-locality in space, which is characterized
by the so-called $k$-mass\cite{mahaux}
\begin{equation}
\frac{m_k(k)}{m} = \left[1 + \frac{m}{k}\frac{\partial\Sigma(k,\omega)}
{\partial k}\right]^{-1}\,,\label{eq:kmass}
\end{equation}
or due to a non-locality in time, which is expressed in terms of the $E$-mass
\begin{equation}
\frac{m_E(\omega)}{m} = \left[1 - \frac{\partial\Sigma(k,\omega)}
{\partial \omega}\right]\,,\label{eq:emass}
\end{equation}
so that the total effective mass is given by
$$
\frac{m^*(k)}{m} = \frac{m_k(k)}{m}\frac{m_E(\omega=\varepsilon_k)}{m}\,.
$$

\begin{figure}
\begin{center}
\center\includegraphics[scale=0.45]{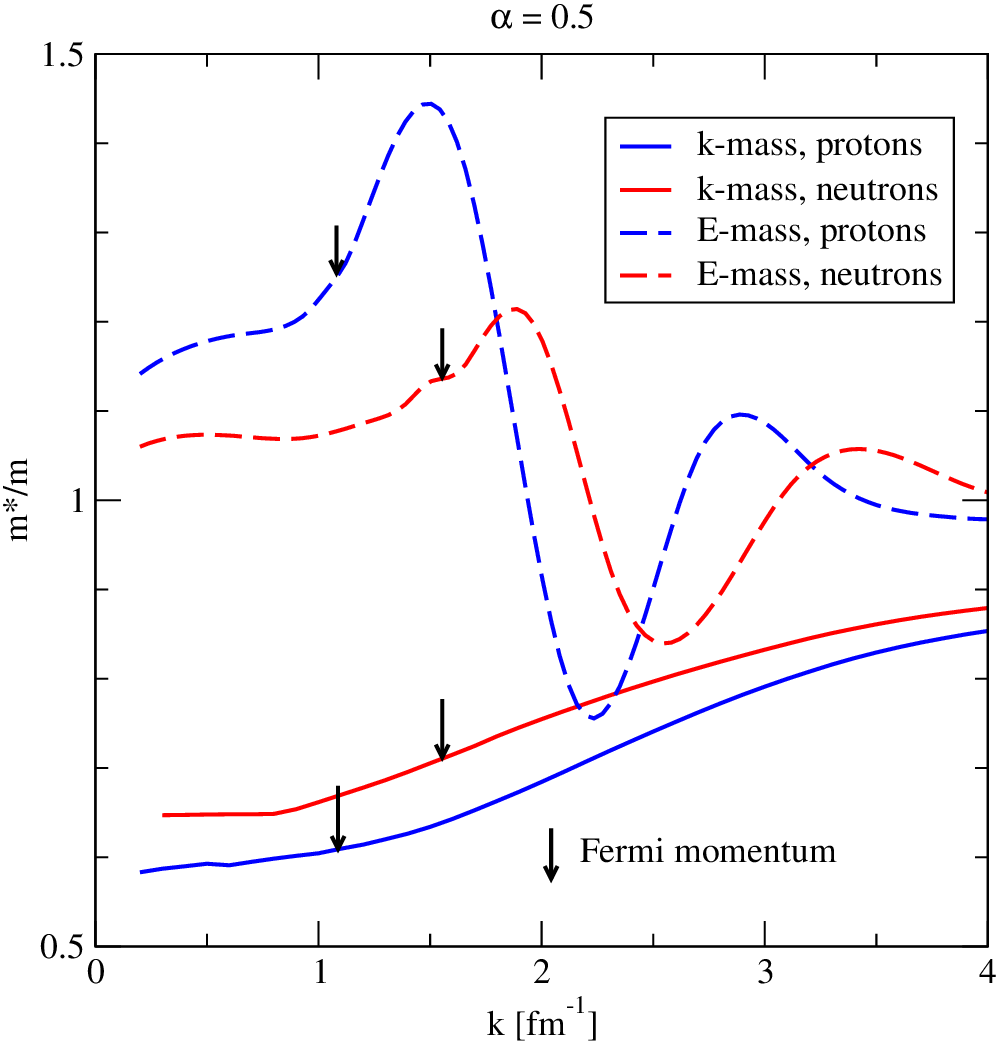}
\end{center}
\caption{\label{fig:epsi2}(Color online) Effective $k$-masses (solid
lines) and $E$-masses (dashed lines) derived from the BHF self-energy. 
Results were obtained using the CD-Bonn interaction at density $\rho$ = 
0.17 fm$^{-1}$ for an asymmetry $\alpha$=0.5.}
\end{figure}

Fig.~\ref{fig:epsi2} shows typical examples for the effective $k$-mass (solid
lines) and the $E$-mass (dashed lines) derived from the BHF self-energy for 
protons and neutrons at asymmetry $\alpha$ = 0.5 using the CD-Bonn interaction. 
The effective $k$-mass shows a rather smooth dependence on the momentum $k$ with
typical values ranging from 0.6 m up to 0.9 m for larger $k$. The effective $k$
mass is hardly effected by the particle-particle ladder
contributions\cite{khaga1}. It is larger for the neutrons than for the protons.     
The effective $E$-masses on the other hand exhibit a strong momentum or energy
dependence (note that the energy variable of the self-energy is related to the
momentum by the self-consistency requirement (\ref{eq:bhf1})). It reaches a
maximum just above the Fermi momentum of the nucleon type under consideration.
These maximal values for the $E$-mass lead to an effective mass $m^*$ close to 
1 at momenta around the Fermi momentum, which corresponds to the characteristic 
momentum dependence of the self-energy displayed in Fig.~\ref{fig:epsi2}.

The energy dependence of the BHF self-energy is due to the particle-particle
ladder terms in the $G$-matrix. The values for the effective $E$-mass are larger
for the protons than for the neutrons (at positive values for the asymmetry
parameter), which is due the fact that particle-particle correlations are of
particular importance for the proton-neutron interaction. The larger $E$-mass
combined with a smaller $k$-mass for the protons leads to effective masses,
which similar for protons and neutrons\cite{elga1,engvik,samaruc}.

\begin{figure}
\begin{center}
\center\includegraphics[scale=0.45]{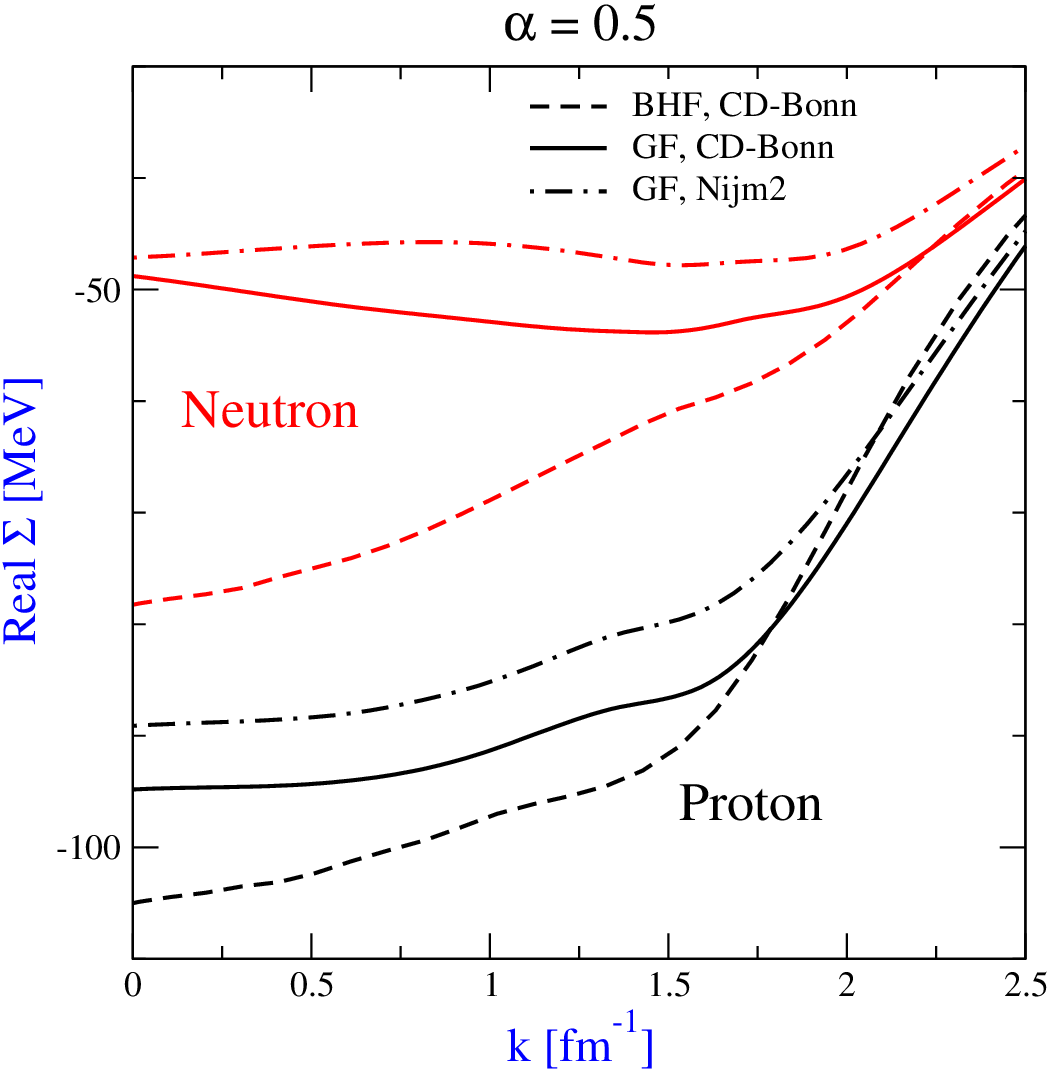}
\end{center}
\caption{\label{fig:qepsi1}(Color online) Single-particle potential determined
from the BHF single-particle energy (\protect{\ref{eq:bhf1}}) and from the 
quasiparticle energy of the Green's function (GF) approach of
(\protect{\ref{eq:qpbhf}}).  Results were obtained using the CD-Bonn and the
Nijm2 interaction at  density $\rho$ = 0.17 fm$^{-1}$ for an asymmetry
$\alpha$=0.5.}
\end{figure}

In the next step we consider the quasi-particle energies defined in
(\ref{eq:qpbhf})  within the Green's function approach (GF). In order to enhance
the momentum dependence, we subtract the contribution for the kinetic energy and
compare in Fig.~\ref{fig:qepsi1} the resulting values for the self-energy with
those determined from the BHF approach. 

The hole-hole ladder contribution leads to less attractive single-particle
energies in the GF as compared to the BHF approach. The difference tends to be
larger for the neutrons than for the protons (at positive values for the
asymmetry parameter $\alpha$) and decreases with increasing momentum. As a
consequence we find that repulsive effect of the hole-hole ladder term is
essentially identical for protons and neutrons at the corresponding Fermi
momenta. 

\begin{figure}
\begin{center}
\center\includegraphics[scale=0.45]{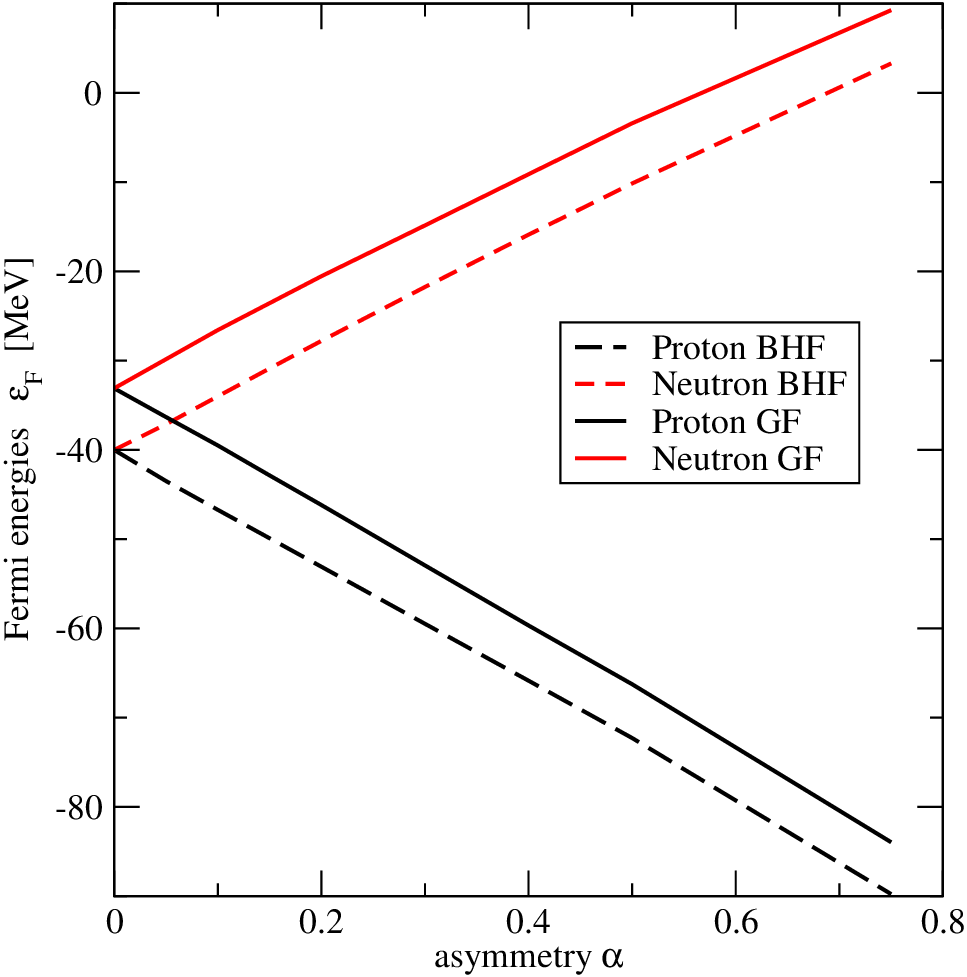}
\end{center}
\caption{\label{fig:epsf}(Color online) Fermi energies for protons and
neutrons derived 
from the BHF single-particle energy (\protect{\ref{eq:bhf1}}) and from the 
quasiparticle energy of the Green's function (GF) approach of
(\protect{\ref{eq:qpbhf}}).  Results were obtained using the CD-Bonn  at  
density $\rho$ = 0.17 fm$^{-1}$. Results are displayed as a function of the
asymmetry parameter $\alpha$.}
\end{figure}

This can be seen from Fig.~\ref{fig:epsf}, which displays the Fermi energies
derived from the BHF and GF single-particle energies for protons and neutrons.
These Fermi energies for protons (neutrons) decrease (increase) linearly with
the asymmetry parameter $\alpha$. The slope is identical for BHF and GF
approach and essentially the same for CD-Bonn and Nijm2 interaction. The only
difference is that the GF Fermi energies are shifted by a value of around 7
MeV. This implies that the conditions for $\beta$-equilibrium are not affected
by the inclusion of the hole-hole scattering terms.    

\begin{figure}
\begin{center}
\center\includegraphics[scale=0.45]{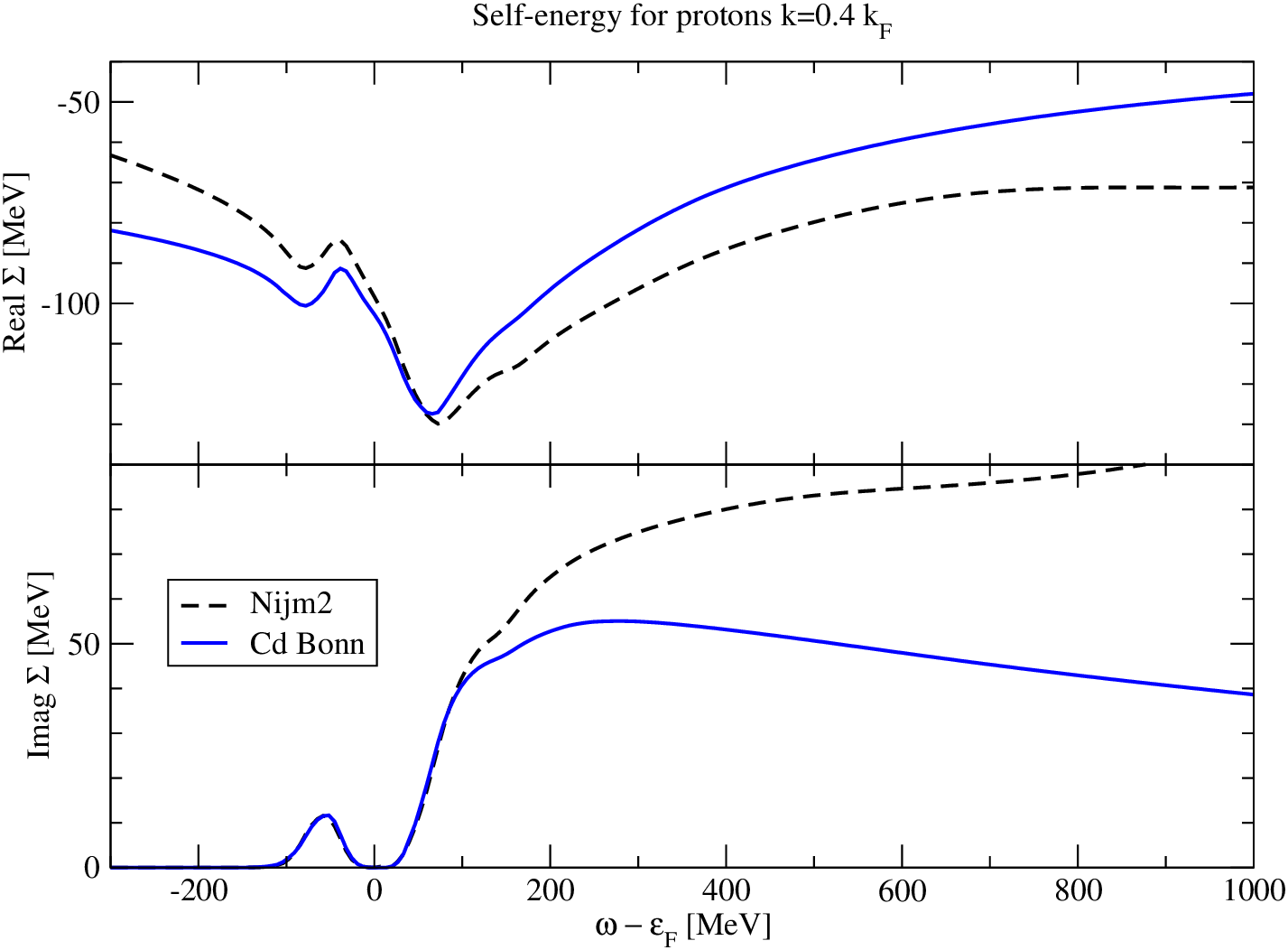}
\end{center}
\caption{\label{fig:qselfp}(Color online) Real (upper panel) and the absolute
value of the imaginary part
(lower panel) of the self-energy for protons with $k$=0.4 $k_{F\pi}$ as a
function of energy $\omega$. The self-energy has been evaluated in the Green's
function approach including $\Sigma^{BHF}$ and $\Delta \Sigma^{2h1p}$ for 
asymmetric nuclear matter at  density $\rho$ = 0.17 fm$^{-1}$ for an asymmetry
$\alpha$=0.5 using the CD-Bonn (solid line) and the Nijm2 (dashed line)
interaction.}
\end{figure}

\begin{figure}
\begin{center}
\center\includegraphics[scale=0.45]{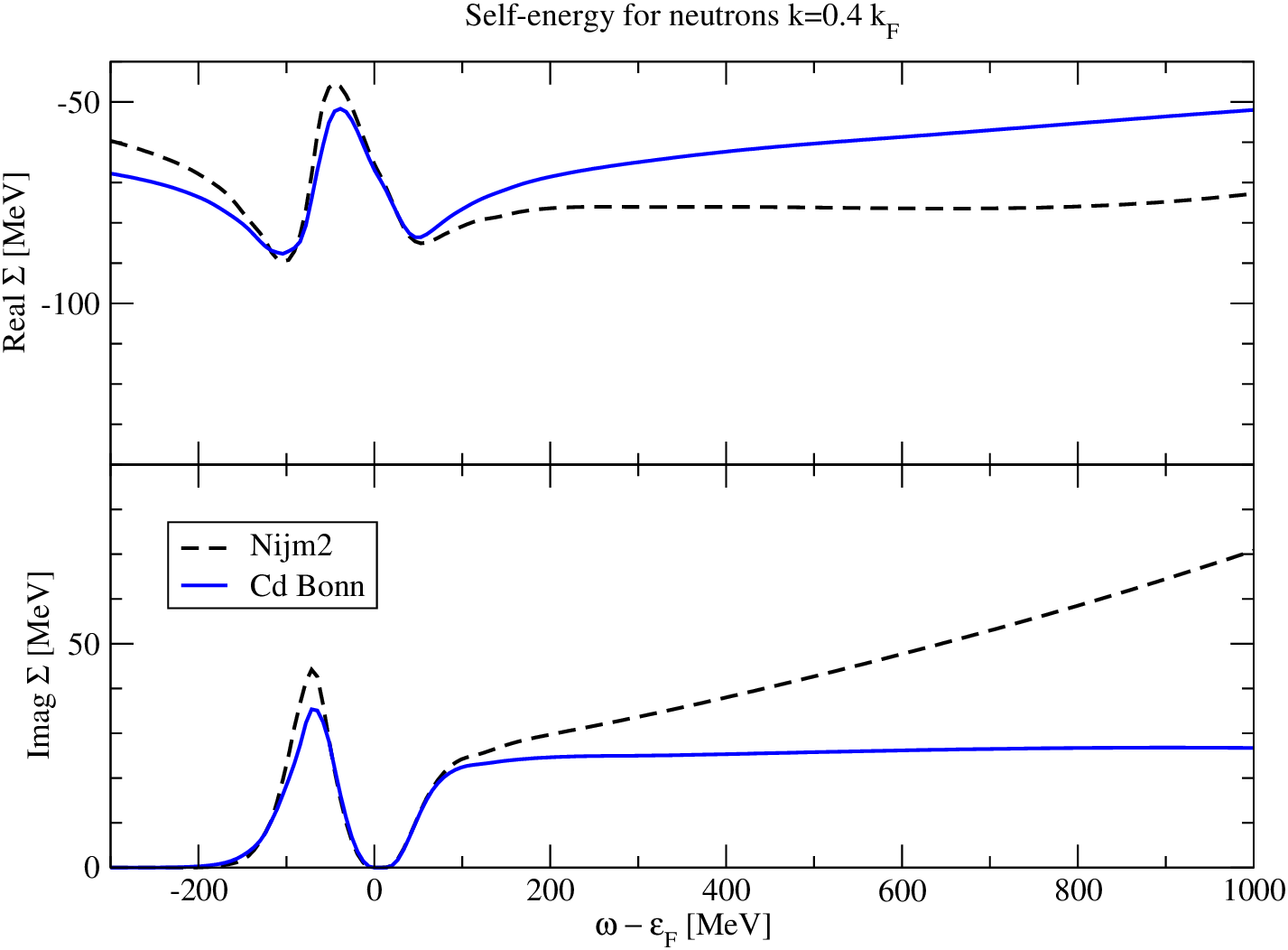}
\end{center}
\caption{\label{fig:qselfn}(Color online) The self-energy for neutrons with 
$k$=0.4 $k_{F\nu}$ as a function of energy $\omega$. Further details see
caption of Fig.~\protect{\ref{fig:qselfp}}}
\end{figure}

In order to investigate the features of the nucleon self-energy more in detail 
we display in Figs.~\ref{fig:qselfp} and \ref{fig:qselfn} the self-energies for
protons and neutrons in asymmetric nuclear matter ($\alpha =0.5$) using the GF 
approach for the CD-Bonn and the Nijm2 interaction. The lower panels of these
figures show the absolute value of the imaginary part of these self-energies
for a fixed momentum as a function of the energy variable $\omega$. The
contributions at energies below the Fermi energy $\varepsilon_F$ originate form
the $2h1p$ part of the self-energy, while those at energies above
$\varepsilon_F$ are due to the BHF part. It is rather obvious that the energy 
integrated $2h1p$ contribution is significantly smaller than the BHF part of the
self-energy. This reflects the fact that for realistic NN interactions and
nuclear densities around the saturation density the particle-particle ladder
contributions are much larger than those of the hole-hole type. This is again a
justification of the hole-line expansion or our perturbative treatment of the
$2h1p$ contribution. 

One also sees rather clearly that the imaginary part calculated from the Nijm2
interaction is larger than the corresponding one determined for the CD-Bonn
interaction. This implies that the local Nijm2 interaction yields larger
scattering amplitudes to high-lying particle-particle states than the non-local
CD-Bonn interaction. The Nijm2 interaction is stiffer than the CD-Bonn
interaction. 

The real part of the self-energy originates from the energy-independent
Hartree-Fock contribution plus the energy-dependent corrections, which are
related to the imaginary part of the self-energy by a dispersion relation.
The gross-structure of the energy-dependence of the real part of $\Sigma$ is due
to the BHF contribution. The larger imaginary part obtained with the Nijm2
interaction leads to a more negative slope of the self-energy in the region
$\omega -\varepsilon_F < -100 MeV$. The fine-structure in the real part of
$\Sigma$ at energies $-100 MeV < \omega -\varepsilon_F < 50 MeV$ has its origin
in the $2h1p$ term and is absent in the BHF approximation. 

Note, that for the asymmetries considered, the $2h1p$ contribution is
significantly larger for the neutron- than for the proton-self-energy. This
indicates, that the imaginary part in $\Sigma^{2h1p}$ increases with the
density of the kind of nucleon under consideration.  On the other hand, the
imaginary part  of the BHF contribution is larger for the protons than for the
neutrons. This is related to the fact that the particle-particle correlations
described by the BHF term are to a large extent due to the strong tensor force
in the proton-neutron interaction. Since the neutron density is increasing with
the asymmetry one can expect a larger imaginary part in $\Sigma^{BHF}$ for the
protons as compared to the neutrons. 

\begin{figure}
\begin{center}
\center\includegraphics[scale=0.45]{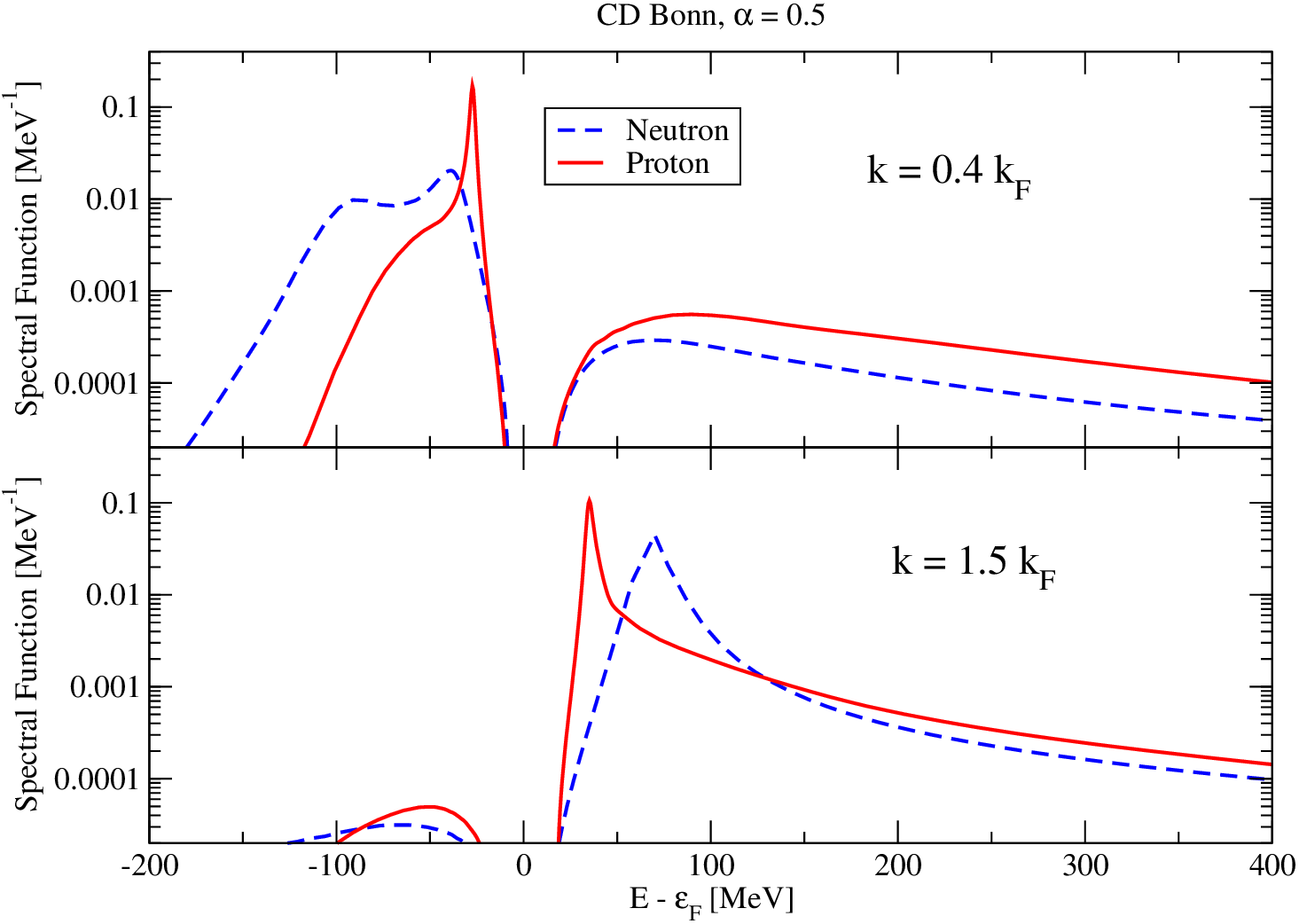}
\end{center}
\caption{\label{fig:spec1}(Color online) Spectral functions for nucleons with 
$k$=0.4 $k_{F}$ in the upper part and $k$=1.5 $k_{F}$ in the lower panel as a 
function of energy $\omega$. Further details see
caption of Fig.~\protect{\ref{fig:qselfp}}}
\end{figure}

The upper part of Fig.~\ref{fig:spec1} displays the spectral functions for
protons and neutrons, considering the same asymmetry and momenta as in 
Figs.~\ref{fig:qselfp} and \ref{fig:qselfn}. While the spectral function for
the proton exhibits a pronounced quasiparticle peak, the spectral function for
the neutron shows a rather broad distribution. This is partly due to the fact
that with this choice of the momentum ($k_\tau$ = 0.4 $k_{F\tau}$) the
quasiparticle  energy minus the corresponding Fermi energy is slightly lower
for the neutrons than for the protons. The main reason for this broader
distribution of the hole strength, however, is the larger imaginary part in
$\Delta \Sigma^{2h1p}$ for the neutrons as compared to the protons, as we
discussed above. 

The larger imaginary part in $\Sigma^{BHF}$ for the protons, on the other hand,
leads to larger values for the proton spectral functions at positive energies 
than for the neutrons, if we consider nucleons with momenta $k < k_F$.  
Consequently, the spectral strength for nucleons with $k>k_F$ at energies below
$\varepsilon_F$ is larger for protons than for the neutrons (see lower panel of
Fig.~\ref{fig:spec1}).

\begin{figure}
\begin{center}
\center\includegraphics[scale=0.45]{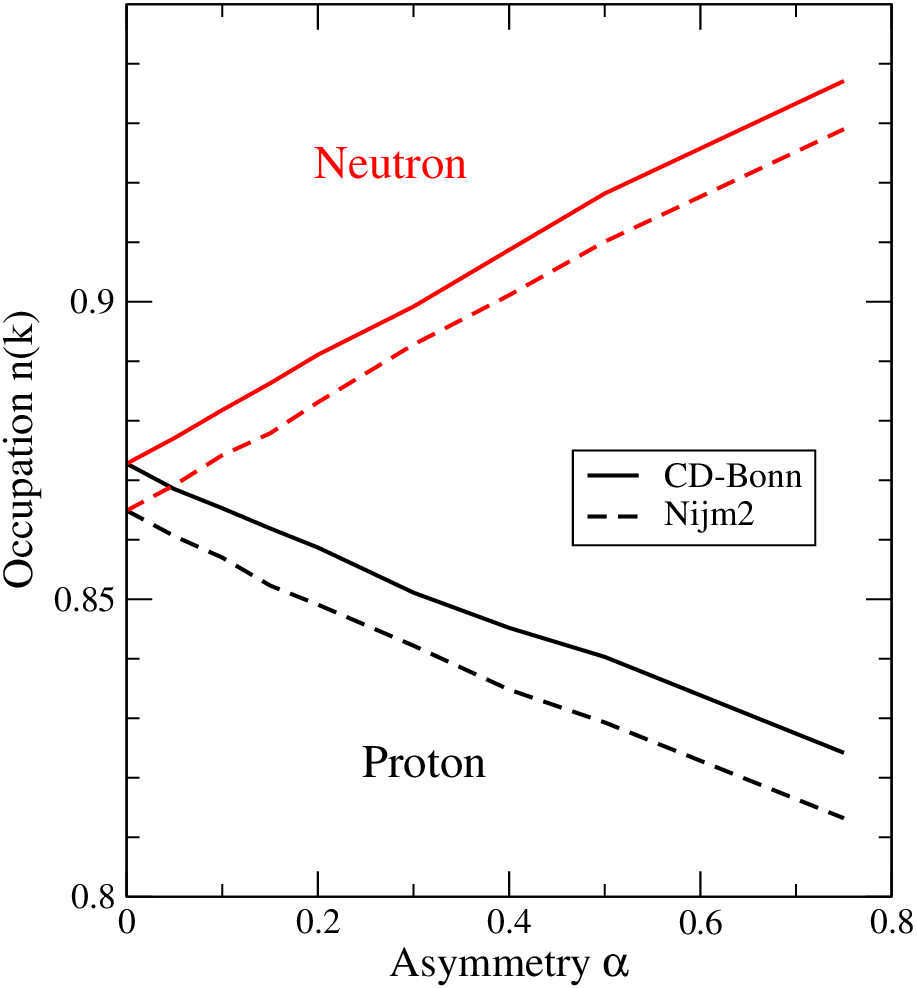}
\end{center}
\caption{\label{fig:nofk}(Color online) Occupation probabilities $n(k)$
according Eq.(\protect{\ref{eq:n(k)}}) averaged over momenta $k < k_F$.}
\end{figure}

This means that the occupation numbers, $n_{\tau}(k)$ (see
Eq.(\ref{eq:n(k)})),  for nucleons with momenta $k$ below the corresponding
Fermi momentum are, at positive values of $\alpha$, smaller for  protons than for
neutrons. Because of the strong proton-neutron interaction, an increase of the
neutron density with increasing asymmetry $\alpha$ yields a larger depletion of
the proton hole-states than of the neutron hole states. This is shown in
Fig.~\ref{fig:nofk}, where occupation probabilities for protons and neutrons,
averaged over all momenta  below the corresponding Fermi momenta, are presented
as a function of asymmetry. One observes an almost linear dependence of the
occupation probabilities as a function of asymmetry. The occupation probabilities
are slightly larger for the CD-Bonn interaction as compared to the stiffer Nijm2
interaction.  

The spectral distribution of hole-strength can be explored in nucleon knock-out
experiments on neutron rich nuclei. Our results would predict that the proton
distribution functions should exhibit more pronounced quasi-particle peaks
in states, which are occupied in the independent particle model, than the
neutrons. Nevertheless, the energy integrated strength in these partial waves,
the occupation numbers, should be larger for the neutrons.

Finally a few words about the total energy per nucleon as a function of
asymmetry. The calculated energies are rather well approximated by a function of
the form
\begin{equation}
\frac{E}{A} = E_0 + a_S \alpha^2\,,
\end{equation}
with the coefficient $a_S$ for the symmetry energy. Results for $E_0$ the
binding energy per nucleon of symmetric nuclear matter at saturation density
and the asymmetry coefficient are listed in  table \ref{tab:ener} and compared
to the empirical values. Note, however, that for these calculations we do not 
determine the saturation density but just calculate at the empirical value of
the saturation density $\rho_0$. 

As we discussed already before, the Nijm2 interaction is a stiffer interaction
and predicts less binding energy per nucleon than the softer CD-Bonn potential. 
The inclusion of 2h1p terms in the self-energy leads to smaller binding energies
per nucleon. The effect is small in the case of the CD-Bonn interaction but
considerably larger for the Nijm2 model. The inclusion of the 2h1p term has only
little effect on the symmetry energy. Also the influence of the interaction is
not very significant.

\begin{table}[ht]
\begin{center}
\begin{tabular}{cc|rr}
&& $E_0$ [MeV] & $a_S$ [MeV]\\
\hline
CD-Bonn & BHF & -18.8 & 31.5 \\
& GF & -18.0 & 33.3 \\
\hline
Nijm2 & BHF & -16.7 & 29.3 \\
& GF & -12.3 & 34.0\\
\hline
& Exp. & -15.7 & 30$\pm$ 2\\
\end{tabular}
\end{center}
\caption{\label{tab:ener} Energy per nucleon, $E_0$ and symmetry energie, $a_S$
determined from BHF and GF calculations using the CD-Bonn and the Nijm2
interaction}
\end{table}

\section{Conclusions}
An extension of the Brueckner Hartree Fock (BHF) approximation to a
self-consistent Green's function method (GF) has been used to study properties
of asymmetric nuclear matter. We investigate in particular the spectral
distribution of single-particle strength for such asymmetric systems. 

It is
observed that the imaginary part of the self-energy at energies below the Fermi
energy increases with the density of the kind of nucleons under consideration.
Since this part is responsible for the spreading of the hole strength, we obtain
more pronounced quasiparticle states for protons than for neutrons in neutron-rich 
matter. 

On the other hand, the occupation probabilities of single-particle states,
which are occupied in the independent particle model, is significantly smaller
for protons than for neutrons in such neutron-rich systems. This can be traced
back to the strong components of proton-neutron interaction, which is the main
source for the imaginary part of the self-energy at momenta below the Fermi
momentum and energies above $\varepsilon_F$. 

The influence of the 2h1p contributions to the self-energy on total energy,
symmetry energy and Fermi energies is not very significant for the soft CD-Bonn
interaction. It leads to considerable effects for the local Nijm2 interaction:
reducing the calculated energy and increasing the coefficient for the symmetry 
energy.  Therefore a more systematic study of asymmetric nuclear systems should
be useful, which goes beyond the present approach and treats the hole-hole
scattering terms in a non-perturbative way\cite{bozek,fricn,bozekn}.         
 
Various discussions with Tobias Frick are gratefully acknowledged.
This work has been supported by the European Graduate School ``Hadron in
Vacuum, Nuclei and Stars'' (Basel - T\"ubingen).

\end{document}